\documentclass{article}
\usepackage{amsfonts}
\usepackage{amssymb}
\usepackage{amsmath}
\usepackage{graphicx}
\usepackage{float}
\usepackage[top=0.75in, bottom=0.75in, left=1.0in, right=1.0in]{geometry}

\begin{document}

\title{Physical interpretation of generalized two-mode squeezing operator
revealed by virtue of the transformation of entangled state representation%
\thanks{%
Work supported by the National Natural Science Foundation of China under
grant (No: 61203061, 61403362, 61374091, 61473199 and 11175113) }}
\author{Gao Fang$^{1}$, Wang Yao-xiong$^{1}$, Fan Hong-yi$^{1,2}$, and Tang
Xu-bing$^{1,3\dag }$ \\
$^{1}$Institute of Intelligent Machines, Chinese Academy of Sciences,\\
Hefei 230031, China\\
$^{2}$Department of Automation, University of Science \& Technology\\
of China, Hefei 230027, China\\
$^{3}$School of Mathematics \& Physics Science and Engineering, \\
Anhui University of Technology, Ma'anshan 243032, China\\
$^{\dag }$ttxxbb@ahut.edu.cn}
\maketitle

\begin{abstract}
By virtue of the integration method within $\mathfrak{P}$-ordered product of
operators and the property of entangled state representation, we reveal new
physical interpretation of the generalized two-mode squeezing operator
(GTSO), and find it be decomposed as the product of free-space propagation
operator, single-mode and two-mode squeezing operators, as well as thin lens
transformation operator. This docomposition is useful to design of opticl
devices for generating various squeezed states of light. Transformation of
entangled state representation induced by GTSO is emphasized.
\end{abstract}


\section{Introduction}

In recent decades, nonclassical optical field (such as squeezed light field
\cite{Loudon_1987_jmo} and non-Gussian optical field \cite%
{Lvovsky_2002_prl,Nielsen_2006_prl,Bartley_2012_pra}) has been received much
attention by physicists as it can be applied to quantum communication
protocols \cite{Enk_2001_pra,Pan_2012_rmp}, metrology \cite%
{Gilchrist_2004_jopb,Xiang_2011_np,Xiang_2013_sr}, cloning \cite%
{Cerf_2005_prl}, quantum computation \cite{Ralph_2003_pra,Lund_2008_prl},
\cite{Wang_2010_pra,Ukai_2011_prl}testing of quantum theory \cite%
{Kim_2008_prl} and high precision quantum measurement, due to the fact that
quadrature measurement of a squeezed state can be less that that of a
coherent state, \cite{Gilchrist_2004_jopb,Xiang_2011_np,Xiang_2013_sr}.
Based on the squeezed sideband machanism, the squeezed light can also be
applied in quantum teleportation \cite{Furusawa_1998_science}. Moreover,
squeezed optical field has proven itself to be quite adapt to quantum key
distribution \cite{Su_2009_epl}, quantum swapping\ \cite{Jia_2004_prl},
controlled quantum dense encoding \cite{Jing_2003_prl}, and quantum phase
tracking \cite{Yonezawa_2012_science}. Theoretical analysis yields the
general conclusion that entangled resource applied in continuous variable
quantum communication is multi-mode squeezed light field. Experimentally,
signal mode and idle mode of the output of a non-degenerate parametric
amplifier constitutes a two-mode squeezed state.

Theoretically, the two-mode squeezing operator $S_{2}$ can be viewed as the
quatum image of mapping of classical scale transformation $\eta \rightarrow
\eta /\mu $ in the bipartite entangled state representation, and has a
concise expression \cite{Fan_2008_ap}%
\begin{equation}
S_{2}=\int \frac{d^{2}\eta }{\pi \mu }\left\vert \eta /\mu \right\rangle
\left\langle \eta \right\vert =e^{\lambda \left( a_{1}^{\dagger
}a_{2}^{\dagger }-a_{1}a_{2}\right) },  \label{squeezing}
\end{equation}%
where $\mu =e^{\lambda }$ is squeezing parameter, $\left\vert \eta
\right\rangle $ \cite{Fan_2003_jopb} is defined as
\begin{equation*}
\left\vert \eta \right\rangle =\exp \left[ -\frac{\left\vert \eta
\right\vert ^{2}}{2}+\eta a_{1}^{\dagger }-\eta ^{\ast }a_{2}^{\dagger
}+a_{1}^{\dagger }a_{2}^{\dagger }\right] \left\vert 00\right\rangle ,\text{
}\eta =\eta _{1}+i\eta _{2},
\end{equation*}%
which is also named Einstein-Podolsky-Rosen (EPR) entangled states,
satisfying eigenvector equations%
\begin{equation*}
\left( Q_{1}-Q_{2}\right) \left\vert \eta \right\rangle =\sqrt{2}\eta
_{1}\left\vert \eta \right\rangle \text{,\ \ }\left( P_{1}+P_{2}\right)
\left\vert \eta \right\rangle =\sqrt{2}\eta _{2}\left\vert \eta \right\rangle
\end{equation*}%
where $Q_{i}=\left( a_{i}+a_{i}^{\dagger }\right) /\sqrt{2}$ and $%
P_{i}=i\left( a_{i}^{\dagger }-a_{i}\right) /\sqrt{2}$ are a pair of
conjugate variables with $\left[ Q_{i},P_{j}\right] =i\delta _{ij},$ $\hbar
=1.$ $S_{2}$ makes $\frac{Q_{1}+Q_{2}}{\sqrt{2}}$ and $\frac{P_{1}+P_{2}}{%
\sqrt{2}}$ as scaling transformation%
\begin{equation}
S_{2}\frac{Q_{1}+Q_{2}}{\sqrt{2}}S_{2}^{-1}=\mu \frac{Q_{1}+Q_{2}}{\sqrt{2}},%
\text{ \ \ }S_{2}\frac{P_{1}+P_{2}}{\sqrt{2}}S_{2}^{-1}=\frac{P_{1}+P_{2}}{%
\sqrt{2}\mu }.  \label{sq_tran}
\end{equation}

An interesting question thus naturally arises: Is there a generalized
two-mode squeezing operator, denoted by $F_{2},$ which is responsible for
generating the following transformations%
\begin{eqnarray}
F_{2}\left( Q_{1}+Q_{2}\right) F_{2}^{-1} &=&A\left( Q_{1}+Q_{2}\right)
+C\left( P_{1}+P_{2}\right)  \notag \\
F_{2}\left( P_{1}+P_{2}\right) F_{2}^{-1} &=&D\left( P_{1}+P_{2}\right)
+B\left( Q_{1}+Q_{2}\right)  \label{transform}
\end{eqnarray}%
in which $AD-BC=1,$ and correspondingly,%
\begin{eqnarray}
F_{2}\left( P_{1}-P_{2}\right) F_{2}^{-1} &=&A\left( P_{1}-P_{2}\right)
-C\left( Q_{1}-Q_{2}\right)  \notag \\
F_{2}\left( Q_{1}-Q_{2}\right) F_{2}^{-1} &=&D\left( Q_{1}-Q_{2}\right)
-B\left( P_{1}-P_{2}\right) ,  \label{trans}
\end{eqnarray}%
and what is the corresponding squeezed sate? In this work, in order to
derive the concrete expression of $F_{2}$, in Sec. II we shall discuss
transformation of the entangled state representation induced by $F_{2}.$ In
Sec. III, by virtue of the integration method within $\mathfrak{P}$-ordered
product of operators (which means that all momentum operators are arranged
on the left of all coordinate operators) we shall show that $F_{2}$ can be
decomposed as a product of free-space propagation operator, single-mode and
two-mode squeezing operator, and thin lens transformation operator. In so
doing, a new physical interpretation of the generalized two-mode squeezing
operator is presented. This docomposition is useful to design of opticl
devices for generating various squeezed states of light. In Sec. IV, we
shall examine the Lie algebra structure of $F_{2}$. In the whole text, we
shall make full use of properties of the entangled state representation.

\section{Representation transformation caused by $F_{2}$}

By introducing another bipartite entangled states $\left\vert \xi
\right\rangle $ in Fock space \cite{Fan_2008_ap}%
\begin{equation}
\left\vert \xi \right\rangle =\exp \left[ -\frac{\left\vert \xi \right\vert
^{2}}{2}+\xi a_{1}^{\dagger }+\xi ^{\ast }a_{2}^{\dagger }-a_{1}^{\dagger
}a_{2}^{\dagger }\left\vert 00\right\rangle \right] ,\text{ }\xi =\xi
_{1}+i\xi _{2},  \label{conjugate}
\end{equation}%
which is the eigenvector of $\left( P_{1}-P_{2}\right) $ and $\left(
Q_{1}+Q_{2}\right) $%
\begin{equation}
\left( Q_{1}+Q_{2}\right) \left\vert \xi \right\rangle =\sqrt{2}\xi
_{1}\left\vert \xi \right\rangle ,\ \left( P_{1}-P_{2}\right) \left\vert \xi
\right\rangle =\sqrt{2}\xi _{2}\left\vert \xi \right\rangle ,  \label{vector}
\end{equation}%
we see the inner product
\begin{equation}
\left\langle \xi \right\vert \left. \eta \right\rangle =\frac{1}{2}%
e^{i\left( \eta _{2}\xi _{1}-\eta _{1}\xi _{2}\right) }=\frac{1}{2}e^{\left(
\eta \xi ^{\ast }-\xi \eta ^{\ast }\right) /2},  \label{added}
\end{equation}%
so $\left\vert \xi \right\rangle $ is the conjugate state to $\left\vert
\eta \right\rangle .$ Since
\begin{equation*}
\left[ D\left( Q_{1}-Q_{2}\right) -B\left( P_{1}-P_{2}\right) ,D\left(
P_{1}+P_{2}\right) +B\left( Q_{1}+Q_{2}\right) \right] =0
\end{equation*}%
\ they should possess common eigenvector, denoted as $\left\vert \eta
\right\rangle _{D,B}$, obeying
\begin{eqnarray}
\left[ D\left( Q_{1}-Q_{2}\right) -B\left( P_{1}-P_{2}\right) \right]
\left\vert \eta \right\rangle _{D,B} &=&\sqrt{2}\eta _{1}\left\vert \eta
\right\rangle _{D,B},  \label{13.1} \\
\left[ B\left( Q_{1}+Q_{2}\right) +D\left( P_{1}+P_{2}\right) \right]
\left\vert \eta \right\rangle _{D,B} &=&\sqrt{2}\eta _{2}\left\vert \eta
\right\rangle _{D,B}.  \label{13.2}
\end{eqnarray}%
From Eq. (\ref{conjugate}), we know%
\begin{equation}
\left\langle \xi \right\vert \left( Q_{1}-Q_{2}\right) =i\sqrt{2}\frac{%
\partial }{\partial \xi _{2}}\left\langle \xi \right\vert ,\text{ \ \ }%
\left\langle \xi \right\vert \left( P_{1}+P_{2}\right) =-i\sqrt{2}\frac{%
\partial }{\partial \xi _{1}}\left\langle \xi \right\vert .  \label{14}
\end{equation}%
Thus from (\ref{13.1}) and (\ref{14}), we have%
\begin{equation}
\left\langle \xi \right\vert \left[ D\left( Q_{1}-Q_{2}\right) -B\left(
P_{1}+P_{2}\right) \right] \left\vert \eta \right\rangle _{D,B}=\sqrt{2}\eta
_{1}\left\langle \xi \right. \left\vert \eta \right\rangle _{D,B}=\sqrt{2}%
\left( Di\frac{\partial }{\partial \xi _{2}}-B\xi _{2}\right) \left\langle
\xi \right. \left\vert \eta \right\rangle _{D,B},  \label{15}
\end{equation}%
its solution is
\begin{equation}
\left\langle \xi \right. \left\vert \eta \right\rangle _{D,B}=\mathfrak{C}%
_{1}\left( \xi _{1},\eta \right) \exp \left[ -\frac{i}{D}\left( \xi _{2}\eta
_{1}+\frac{B\xi _{2}^{2}}{2}\right) \right] ,  \label{solution1}
\end{equation}%
where $\mathfrak{C}_{1}\left( \xi _{1},\eta \right) $ is an integral
constant. In the same way, from (\ref{13.2}) we can obtain
\begin{equation*}
\left\langle \xi \right\vert \left[ B\left( Q_{1}+Q_{2}\right) +D\left(
P_{1}+P_{2}\right) \right] \left\vert \eta \right\rangle _{D,B}=\sqrt{2}\eta
_{2}\left\vert \eta \right\rangle _{D,B}=\sqrt{2}\left( B\xi _{1}-Di\frac{%
\partial }{\partial \xi _{1}}\right) \left\langle \xi \right. \left\vert
\eta \right\rangle _{D,B}
\end{equation*}%
with the solution
\begin{equation}
\left\langle \xi \right. \left\vert \eta \right\rangle _{D,B}=\mathfrak{C}%
_{2}\left( \xi _{2},\eta \right) \exp \left[ \frac{i}{D}\left( \xi _{1}\eta
_{2}-\frac{B\xi _{1}^{2}}{2}\right) \right] .  \label{solution2}
\end{equation}%
Eqs. (\ref{solution1}) and (\ref{solution2}) tell us%
\begin{equation}
\left\langle \xi \right. \left\vert \eta \right\rangle _{D,B}=\mathfrak{C}%
\left( \eta \right) \exp \left[ \frac{i}{D}\left( \xi _{1}\eta _{2}-\xi
_{2}\eta _{1}-\frac{B|\xi |^{2}}{2}\right) \right] ,  \label{solution3}
\end{equation}%
$\mathfrak{C}\left( \eta \right) $ is an integral constant. Using the
completeness relation of $\left\vert \xi \right\rangle ,$ i.e.$\ \int \frac{%
d^{2}\xi }{\pi }\left\vert \xi \right\rangle \left\langle \xi \right\vert
=1, $ we have%
\begin{align*}
& \left\vert \eta \right\rangle _{D,B}=\int \frac{d^{2}\xi }{\pi }\left\vert
\xi \right\rangle \left\langle \xi \right. \left\vert \eta \right\rangle
_{D,B} \\
& =\mathfrak{C}\left( \eta \right) \frac{2D}{D+iB}\exp \{\frac{1}{D+iB}\left[
\frac{-|\eta |^{2}}{2D}+\eta a_{1}^{\dagger }-\eta ^{\ast }a_{2}^{\dagger
}+\left( D-iB\right) a_{1}^{\dagger }a_{2}^{\dagger }\right] \}\left\vert
00\right\rangle .
\end{align*}%
If setting $\mathfrak{C}\left( \eta \right) =\frac{1}{2D}\exp \left[ \frac{iC%
}{2D}|\eta |^{2}\right] ,$ we obtain%
\begin{equation*}
\left\vert \eta \right\rangle _{D,B}=\frac{1}{D+iB}\exp \left[ -\frac{A-iC}{%
2\left( D+iB\right) }|\eta |^{2}+\frac{\eta a_{1}^{\dagger }}{D+iB}-\frac{%
\eta ^{\ast }a_{2}^{\dagger }}{D+iB}+\frac{D-iB}{D+iB}a_{1}^{\dagger
}a_{2}^{\dagger }\right] ,
\end{equation*}%
which is complete, $\int \frac{d^{2}\eta }{\pi }\left\vert \eta
\right\rangle _{D,B}$ $_{D,B}\left\langle \eta \right\vert =1.$ Consulting (%
\ref{transform}) and (\ref{trans}) and using$\ $orthogonality of the
entangled state representation $\left\vert \eta \right\rangle $, $%
\left\langle \eta \right\vert \left. \eta ^{\prime }\right\rangle =\pi
\delta ^{\left( 2\right) }\left( \eta -\eta ^{\prime }\right) ,$ we know
that $F_{2}$ can be expressed as%
\begin{equation}
F_{2}=\int \frac{d^{2}\eta }{\pi }\left\vert \eta \right\rangle _{D,B\ }%
\text{ }\left\langle \eta \right\vert .  \label{expression}
\end{equation}%
Thus the unitary operator $F_{2}$ can convert $\left\vert \eta \right\rangle
$ into $\left\vert \eta \right\rangle _{D,B},$ (\ref{expression}) showing
the representation transformation caused by $F_{2}.$

\section{Explicit Expression of $F_{2}$ and its decomposition}

Considering the completeness relation $\int \frac{d^{2}\xi }{\pi }\left\vert
\xi \right\rangle \left\langle \xi \right\vert =1$ and noting (\ref{added})
we have
\begin{eqnarray*}
\frac{1}{2}\left\vert \xi \right\rangle \left\langle \eta \right\vert
e^{-\left( \xi \eta ^{\ast }-\eta \xi ^{\ast }\right) /2} &=&\pi ^{2}\delta
\left( \xi _{2}-\frac{P_{1}-P_{2}}{\sqrt{2}}\right) \delta \left( \xi _{1}-%
\frac{Q_{1}+Q_{2}}{\sqrt{2}}\right) \\
&&\times \delta \left( \eta _{2}-\frac{P_{1}+P_{2}}{\sqrt{2}}\right) \delta
\left( \eta _{1}-\frac{Q_{1}-Q_{2}}{\sqrt{2}}\right) .
\end{eqnarray*}%
Then using Eq. (\ref{solution3}) and (\ref{expression}) we derive
\begin{eqnarray}
F_{2} &=&\int \frac{d^{2}\eta }{\pi }\int \frac{d^{2}\xi }{\pi }\left\vert
\xi \right\rangle \left\langle \xi \right\vert \left. \eta \right\rangle
_{D,B\ }\left\langle \eta \right\vert =\int \frac{d^{2}\eta }{\pi }\int
\frac{d^{2}\xi }{\pi }\frac{1}{2}\left\vert \xi \right\rangle \left\langle
\eta \right\vert e^{-\left( \xi \eta ^{\ast }-\eta \xi ^{\ast }\right) /2}
\notag \\
&&\times \frac{1}{D}\exp \left[ i\frac{C|\eta |^{2}-B|\xi |^{2}}{2D}+\frac{%
i\left( \xi _{1}\eta _{2}-\xi _{2}\eta _{1}\right) \left( 1-D\right) }{D}%
\right]  \notag \\
&=&\frac{1}{D}\int d^{2}\eta \int d^{2}\xi \delta \left( \xi _{2}-\frac{%
P_{1}-P_{2}}{\sqrt{2}}\right) \delta \left( \xi _{1}-\frac{Q_{1}+Q_{2}}{%
\sqrt{2}}\right) \delta \left( \eta _{2}-\frac{P_{1}+P_{2}}{\sqrt{2}}\right)
\notag \\
&&\times \delta \left( \eta _{1}-\frac{Q_{1}-Q_{2}}{\sqrt{2}}\right) \exp %
\left[ i\frac{C|\eta |^{2}-B|\xi |^{2}}{2D}+\frac{i\left( \xi _{1}\eta
_{2}-\xi _{2}\eta _{1}\right) \left( 1-D\right) }{D}\right] .  \label{17}
\end{eqnarray}%
Recalling that in Ref. \cite{Fan_2013_sc,r1,r2,r3} the authors has proposed
a new approach for handling $\mathfrak{Q}$-ordering (all $Q$ are on the left
of all $P$) and $\mathfrak{P}$-ordering (all $P$ are on the left of all $Q$%
). By virtue of the integration method within $\mathfrak{P}$-ordered product
of operator we have%
\begin{equation}
\delta \left( \xi _{1}-\frac{Q_{1}+Q_{2}}{\sqrt{2}}\right) \delta \left(
\eta _{2}-\frac{P_{1}+P_{2}}{\sqrt{2}}\right) =\frac{1}{2\pi }\mathfrak{P}%
\left[ e^{i\left( \eta _{2}-\frac{P_{1}+P_{2}}{\sqrt{2}}\right) \left( \xi
_{1}-\frac{Q_{1}+Q_{2}}{\sqrt{2}}\right) }\right] .  \label{172}
\end{equation}%
Thus the last equation of Eq. (\ref{17}) reads as%
\begin{align*}
F_{2}& =\frac{1}{2\pi D}\int d^{2}\eta \int d^{2}\xi \delta \left( \xi _{2}-%
\frac{P_{1}-P_{2}}{\sqrt{2}}\right) \mathfrak{P}\left[ e^{i\left( \eta _{2}-%
\frac{P_{1}+P_{2}}{\sqrt{2}}\right) \left( \xi _{1}-\frac{Q_{1}+Q_{2}}{\sqrt{%
2}}\right) }\right] \delta \left( \eta _{1}-\frac{Q_{1}-Q_{2}}{\sqrt{2}}%
\right) \\
& \times \exp \left[ i\frac{C|\eta |^{2}-B|\xi |^{2}}{2D}+\frac{i\left( \xi
_{1}\eta _{2}-\xi _{2}\eta _{1}\right) \left( 1-D\right) }{D}\right] \\
& =\frac{1}{2\pi D}\mathfrak{P}\left\{ \int d\eta _{2}\int d\xi
_{1}e^{i\left( \eta _{2}-\frac{P_{1}+P_{2}}{\sqrt{2}}\right) \left( \xi _{1}-%
\frac{Q_{1}+Q_{2}}{\sqrt{2}}\right) }\exp \left[ i\frac{C\eta _{2}^{2}+C%
\frac{\left( Q_{1}-Q_{2}\right) ^{2}}{2}-B\xi _{1}^{2}-B\frac{\left(
P_{1}-P_{2}\right) ^{2}}{2}}{2D}\right. \right. \\
& +\left. \left. \frac{i\left( \xi _{1}\eta _{2}-\frac{P_{1}-P_{2}}{\sqrt{2}}%
\frac{Q_{1}-Q_{2}}{\sqrt{2}}\right) \left( 1-D\right) }{D}\right] \right\} .
\end{align*}%
Then performing integration over $d\xi _{1}$ and $d\eta _{2},$ we obtain%
\begin{align}
F_{2}& =\sqrt{\frac{1}{AD}}\exp \left[ \frac{iC}{4A}\left(
P_{1}+P_{2}\right) ^{2}-\frac{iB}{4D}\left( P_{1}-P_{2}\right) ^{2}\right]
\notag \\
& \times \mathfrak{P}\exp \left\{ -i\left( P_{1}\ P_{2}\right) \left(
\begin{array}{cc}
\frac{1}{2A}+\frac{1}{2D}-1 & \frac{1}{2A}-\frac{1}{2D} \\
\frac{1}{2A}-\frac{1}{2D} & \frac{1}{2A}+\frac{1}{2D}-1%
\end{array}%
\right) \left(
\begin{array}{c}
Q_{1} \\
Q_{2}%
\end{array}%
\right) \right\}  \label{18} \\
& \times \exp \left[ \frac{iC}{4D}\left( Q_{1}-Q_{2}\right) ^{2}-\frac{iB}{4A%
}\left( Q_{1}+Q_{2}\right) ^{2}\right] .  \notag
\end{align}%
Then using the operator identity \cite{r2}
\begin{equation}
e^{-i\vec{P}\Lambda \vec{Q}}\equiv e^{\left( -iP\right) _{l}\Lambda
_{lk}Q_{k}}=\mathfrak{P}\left[ \exp \{\left( -iP\right) _{l}\left(
e^{\Lambda }-1\right) _{lk}Q_{k}\right] =\mathfrak{P}e^{-i\vec{P}\left(
e^{\Lambda }-1\right) \vec{Q}}  \label{33}
\end{equation}%
and%
\begin{equation*}
\left[ \left( P_{1}Q_{1}+P_{2}Q_{2}\right) ,\left(
P_{1}Q_{2}+P_{2}Q_{1}\right) \right] =0,
\end{equation*}%
we have%
\begin{align}
& \mathfrak{P}\exp \{-i\left( P_{1}\ P_{2}\right) \left(
\begin{array}{cc}
\frac{1}{2A}+\frac{1}{2D}-1 & \frac{1}{2A}-\frac{1}{2D} \\
\frac{1}{2A}-\frac{1}{2D} & \frac{1}{2A}+\frac{1}{2D}-1%
\end{array}%
\right) \left(
\begin{array}{c}
Q_{1} \\
Q_{2}%
\end{array}%
\right) \}  \notag \\
& =\exp \left[ -i\left( P_{1}\ P_{2}\right) \left(
\begin{array}{cc}
-\ln \sqrt{AD} & \ln \sqrt{\frac{D}{A}} \\
\ln \sqrt{\frac{D}{A}} & -\ln \sqrt{AD}%
\end{array}%
\right) \left(
\begin{array}{c}
Q_{1} \\
Q_{2}%
\end{array}%
\right) \right]  \label{19} \\
& =\exp \left[ i\left( P_{1}Q_{1}+P_{2}Q_{2}\right) \ln \sqrt{AD}\right]
\exp \left[ -i\left( P_{1}Q_{2}+P_{2}Q_{1}\right) \ln \sqrt{\frac{D}{A}}%
\right] .  \notag
\end{align}%
Noting
\begin{equation}
\ln \left(
\begin{array}{cc}
\frac{1}{2A}+\frac{1}{2D} & \frac{1}{2A}-\frac{1}{2D} \\
\frac{1}{2A}-\frac{1}{2D} & \frac{1}{2A}+\frac{1}{2D}%
\end{array}%
\right) =\left(
\begin{array}{cc}
-\ln \sqrt{AD} & \ln \sqrt{\frac{D}{A}} \\
\ln \sqrt{\frac{D}{A}} & -\ln \sqrt{AD}%
\end{array}%
\right)  \label{34}
\end{equation}%
and in reference to Eq. (\ref{19}), we consider the following transformation
\begin{eqnarray}
e^{-i\left( P_{1}Q_{2}+P_{2}Q_{1}\right) \ln \sqrt{\frac{D}{A}}}\left(
Q_{1}+Q_{2}\right) e^{i\left( P_{1}Q_{2}+P_{2}Q_{1}\right) \ln \sqrt{\frac{D%
}{A}}} &=&\sqrt{\frac{A}{D}}\left( Q_{1}+Q_{2}\right) ,  \notag \\
e^{-i\left( P_{1}Q_{2}+P_{2}Q_{1}\right) \ln \sqrt{\frac{D}{A}}}\left(
Q_{1}-Q_{2}\right) e^{i\left( P_{1}Q_{2}+P_{2}Q_{1}\right) \ln \sqrt{\frac{D%
}{A}}} &=&\sqrt{\frac{D}{A}}\left( Q_{1}-Q_{2}\right) ,  \label{20} \\
e^{-i\left( P_{1}Q_{2}+P_{2}Q_{1}\right) \ln \sqrt{\frac{D}{A}}}\left(
P_{1}+P_{2}\right) e^{i\left( P_{1}Q_{2}+P_{2}Q_{1}\right) \ln \sqrt{\frac{D%
}{A}}} &=&\sqrt{\frac{D}{A}}\left( P_{1}+P_{2}\right) ,  \notag
\end{eqnarray}%
which indicates $e^{-i\left( P_{1}Q_{2}+P_{2}Q_{1}\right) \ln \sqrt{\frac{D}{%
A}}}$ is a two-mode squeezing operator and
\begin{align}
e^{i\left( P_{1}Q_{1}+P_{2}Q_{2}\right) \ln \sqrt{AD}}\left( Q_{1}\pm
Q_{2}\right) e^{-i\left( P_{1}Q_{1}+P_{2}Q_{2}\right) \ln \sqrt{AD}}& =\sqrt{%
AD}\left( Q_{1}\pm Q_{2}\right) ,  \notag \\
e^{i\left( P_{1}Q_{1}+P_{2}Q_{2}\right) \ln \sqrt{AD}}\left( P_{1}\pm
P_{2}\right) e^{-i\left( P_{1}Q_{1}+P_{2}Q_{2}\right) \ln \sqrt{AD}}& =\frac{%
1}{\sqrt{AD}}\left( P_{1}\pm P_{2}\right) .  \label{21}
\end{align}%
Besides, in reference to Eq. (\ref{18}), we see
\begin{equation*}
\exp \left[ -\frac{iB}{4A}\left( Q_{1}+Q_{2}\right) ^{2}\right] \left(
P_{1}+P_{2}\right) \exp \left[ \frac{iB}{4A}\left( Q_{1}+Q_{2}\right) ^{2}%
\right] =P_{1}+P_{2}+\frac{B}{A}\left( Q_{1}+Q_{2}\right) .
\end{equation*}%
Combining the last three results leads to Eqs. (\ref{transform}) and (\ref%
{trans}). Therefore, from Eqs. (\ref{18}) and (\ref{19}) we conclude that
\begin{align}
F_{2}& =\sqrt{\frac{1}{AD}}\exp \left[ \frac{iC}{4A}\left(
P_{1}+P_{2}\right) ^{2}-\frac{iB}{4D}\left( P_{1}-P_{2}\right) ^{2}\right]
\notag \\
& \times \exp \left[ i\left( P_{1}Q_{1}+P_{2}Q_{2}\right) \ln \sqrt{AD}%
\right] \exp \left[ -i\left( P_{1}Q_{2}+P_{2}Q_{1}\right) \ln \sqrt{\frac{D}{%
A}}\right]  \label{22} \\
& \times \exp \left[ \frac{iC}{4D}\left( Q_{1}-Q_{2}\right) ^{2}-\frac{iB}{4A%
}\left( Q_{1}+Q_{2}\right) ^{2}\right] .  \notag
\end{align}%
is we seeked for. In Eq. (\ref{22}), according to the terminology in matrix
optics theory, the first term denotes a freespace propagation operator, the
second and the third term are respectively the single-mode and two-mode
squeezing operator, and last term is a thin lens transformation operator.
This decomposition is useful to design of opticl devices for generating
various squeezed states of light.

\section{Lie algebra of Equation (\protect\ref{22})}

In order to further explain that Eq. (\ref{22}) is really a new
decomposition, we study its ingredient
\begin{align}
& \sqrt{\frac{1}{AD}}\exp \left[ -\frac{iB}{4D}\left( P_{1}-P_{2}\right) ^{2}%
\right] \exp \left[ i\left( P_{1}Q_{1}+P_{2}Q_{2}\right) \ln \sqrt{AD}\right]
\notag \\
& \times \exp \left[ -i\left( P_{1}Q_{2}+P_{2}Q_{1}\right) \ln \sqrt{\frac{D%
}{A}}\right] \exp \left[ \frac{iC}{4D}\left( Q_{1}-Q_{2}\right) ^{2}\right]
\equiv G.  \label{23}
\end{align}%
We can prove%
\begin{equation}
G=\exp \left\{ \frac{iC}{4A}\left( Q_{1}-Q_{2}\right) ^{2}\right\} \exp %
\left[ i\left( Q_{1}P_{2}+Q_{2}P_{1}\right) \ln A\right] \exp \left\{ \frac{%
-iB}{4A}\left( P_{1}-P_{2}\right) ^{2}\right\}  \label{24}
\end{equation}%
so we can convert Eq. (\ref{22}) into%
\begin{align}
F_{2}& =\exp \left\{ \frac{iC}{4A}\left[ \left( Q_{1}-Q_{2}\right)
^{2}+\left( P_{1}+P_{2}\right) ^{2}\right] \right\} \exp \left[ i\left(
Q_{1}P_{2}+Q_{2}P_{1}\right) \ln A\right]  \label{25} \\
& \times \exp \left\{ \frac{-iB}{4A}\left[ \left( Q_{1}+Q_{2}\right)
^{2}+\left( P_{1}-P_{2}\right) ^{2}\right] \right\} .  \notag
\end{align}%
its merit lies in the obvious SU(1,1) Lie algebra structure
\begin{align}
\left[ \frac{\left( Q_{1}-Q_{2}\right) ^{2}+\left( P_{1}+P_{2}\right) ^{2}}{4%
},\frac{\left( Q_{1}+Q_{2}\right) ^{2}+\left( P_{1}-P_{2}\right) ^{2}}{4}%
\right] & =\frac{-i}{2}\left( Q_{1}P_{2}+Q_{2}P_{1}\right) ,  \notag \\
\left[ \frac{-i}{2}\left( Q_{1}P_{2}+Q_{2}P_{1}\right) ,\frac{\left(
Q_{1}-Q_{2}\right) ^{2}+\left( P_{1}+P_{2}\right) ^{2}}{4}\right] & =\frac{%
\left( Q_{1}-Q_{2}\right) ^{2}+\left( P_{1}+P_{2}\right) ^{2}}{4},
\label{26} \\
\left[ \frac{-i}{2}\left( Q_{1}P_{2}+Q_{2}P_{1}\right) ,\frac{\left(
Q_{1}+Q_{2}\right) ^{2}+\left( P_{1}-P_{2}\right) ^{2}}{4}\right] & =-\frac{%
\left( Q_{1}+Q_{2}\right) ^{2}+\left( P_{1}-P_{2}\right) ^{2}}{4},  \notag
\end{align}%
thus $F_{2}$ embodies SU(1,1) Lie algebra. Now we show that the equivalence
between Eq. (\ref{24}) and Eq. (\ref{23}) can be derived by using the
integration method within $\mathfrak{P}$-ordered product of operators. In
fact, using the entangled state representation $\left\vert \xi \right\rangle
,$ we have
\begin{equation}
\exp \left[ i\left( Q_{1}P_{2}+Q_{2}P_{1}\right) \ln A\right] \left\vert \xi
\right\rangle =\frac{1}{A}\left\vert \xi /A\right\rangle  \label{27}
\end{equation}%
and using the completeness relation of $\left\vert \xi \right\rangle $ and $%
\left\vert \eta \right\rangle ,$ we can see
\begin{align}
& (26)=\exp \left[ \frac{iC}{4A}\left( Q_{1}-Q_{2}\right) ^{2}\right] \int
\frac{d^{2}\eta }{\pi }\left\vert \eta \right\rangle \left\langle \eta
\right\vert \exp \left[ i\left( Q_{1}P_{2}+Q_{2}P_{1}\right) \ln A\right]
\int \frac{d^{2}\xi }{\pi }\left\vert \xi \right\rangle \left\langle \xi
\right\vert \exp \left[ \frac{-iB}{4A}\left( P_{1}-P_{2}\right) ^{2}\right]
\notag \\
& =\int \frac{d^{2}\eta }{\pi }\int \frac{d^{2}\xi }{A\pi }\exp \left( \frac{%
iC\eta _{1}^{2}}{2A}\right) \left[ \frac{1}{2}\left\vert \eta \right\rangle
\left\langle \xi \right\vert e^{\frac{\xi \eta ^{\ast }-\xi ^{\ast }\eta }{2}%
}\right] \exp \left( -\frac{iB\xi _{2}^{2}}{2A}+\frac{\xi \eta ^{\ast }-\xi
^{\ast }\eta }{2}\left( \frac{1}{A}-1\right) \right) ,  \label{28}
\end{align}%
in which
\begin{equation*}
\left\vert \eta \right\rangle \left\langle \xi \right\vert e^{\frac{\xi \eta
^{\ast }-\xi ^{\ast }\eta }{2}}=2\pi ^{2}\delta \left( \eta _{2}-\frac{%
P_{1}+P_{2}}{\sqrt{2}}\right) \delta \left( \eta _{1}-\frac{Q_{1}-Q_{2}}{%
\sqrt{2}}\right) \delta \left( \xi _{2}-\frac{P_{1}-P_{2}}{\sqrt{2}}\right)
\delta \left( \xi _{1}-\frac{Q_{1}+Q_{2}}{\sqrt{2}}\right) ,
\end{equation*}%
it then follows from
\begin{equation}
\delta \left( \eta _{1}-\frac{Q_{1}-Q_{2}}{\sqrt{2}}\right) \delta \left(
\xi _{2}-\frac{P_{1}-P_{2}}{\sqrt{2}}\right) =\frac{1}{2\pi }\mathfrak{P}%
\left[ e^{i\left( \xi _{2}-\frac{P_{1}-P_{2}}{\sqrt{2}}\right) \left( \eta
_{1}-\frac{Q_{1}-Q_{2}}{\sqrt{2}}\right) }\right]  \label{30}
\end{equation}%
and Eq. (\ref{19}) we have%
\begin{align}
(26)& =\frac{1}{2\pi A}\int d^{2}\eta \int d^{2}\xi e^{\frac{iC\eta _{1}^{2}%
}{2A}}e^{\frac{-iB\xi _{2}^{2}}{2A}}\delta \left( \eta _{2}-\frac{P_{1}+P_{2}%
}{\sqrt{2}}\right) \mathfrak{P}\left[ e^{i\left( \xi _{2}-\frac{P_{1}-P_{2}}{%
\sqrt{2}}\right) \left( \eta _{1}-\frac{Q_{1}-Q_{2}}{\sqrt{2}}\right) }%
\right]  \notag \\
& \times \delta \left( \xi _{1}-\frac{Q_{1}+Q_{2}}{\sqrt{2}}\right)
e^{i\left( \eta _{1}\xi _{2}-\eta _{2}\xi _{1}\right) \left( \frac{1}{A}%
-1\right) }  \notag \\
& =\sqrt{\frac{1}{AD}}e^{-\frac{iB}{4D}\left( P_{1}-P_{2}\right) ^{2}}
\label{29} \\
& \times \mathfrak{P}\exp \left\{ -i\left( P_{1}\ P_{2}\right) \left(
\begin{array}{cc}
\frac{1}{2A}+\frac{1}{2D}-1 & \frac{1}{2A}-\frac{1}{2D} \\
\frac{1}{2A}-\frac{1}{2D} & \frac{1}{2A}+\frac{1}{2D}-1%
\end{array}%
\right) \left(
\begin{array}{c}
Q_{1} \\
Q_{2}%
\end{array}%
\right) \right\} e^{\frac{iC}{4D}\left( Q_{1}-Q_{2}\right) ^{2}}  \notag \\
& =(25)  \notag
\end{align}

In summary, By virtue of the integration method within $\mathfrak{P}$%
-ordered product of operators and the property of entangled state
representation, we reveal new physical interpretation of the generalized
two-mode squeezing operator (GTSO), and find it be decomposed as the product
of free-space propagation operator, single-mode and two-mode squeezing
operators, as well as thin lens transformation operator. This docomposition
provides experimantalists with a new scheme for generating various two-mode
squeezed states, which may have potential application in quantum information
and high-precision quantum metrology.

\bigskip

\end{document}